# Mid- and far-infrared localized surface plasmon resonances in chalcogen-hyperdoped silicon

Mao Wang[1,*], Ye Yu[2,‡], Slawomir Prucnal[1], Yonder Berencén[1], Mohd Saif Shaikh[1], Lars Rebohle[1], Muhammad Bilal Khan[1], Vitaly Zviagin[3], René Hübner[1], Alexej Pashkin[1], Artur Erbe[1,4], Yordan M. Georgiev[1,5], Marius Grundmann[3], Manfred Helm[1,6], Robert Kirchner[2,4] and Shengqiang Zhou[1]

[1]Helmholtz-Zentrum Dresden-Rossendorf, Institute of Ion Beam Physics and Materials Research, Bautzner Landstraße 400, 01328 Dresden, Germany

[2]Institute of Semiconductors and Microsystems, Technische Universität Dresden, 01062 Dresden, Germany

[3]Felix-Bloch-Institut für Festkörperphysik, Universität Leipzig, Linnéstraße 5, 04103 Leipzig, Germany

[4]Centre for Advancing Electronics Dresden (CfAED), Technische Universität Dresden, 01062 Dresden, Germany

[5]Institute of Electronics at the Bulgarian Academy of Sciences, 1784 Sofia, Bulgaria

[6]Institut für Angewandte Physik (IAP), Technische Universität Dresden, 01062 Dresden, Germany

**Abstract**

Plasmonic sensing in the infrared region employs the direct interaction of the vibrational fingerprints of molecules with the plasmonic resonances, creating surface-enhanced sensing platforms that are superior than the traditional spectroscopy. However, the standard noble metals used for plasmonic resonances suffer from high radiative losses as well as fabrication challenges, such as tuning the spectral resonance positions into mid- to far-infrared regions, and the compatibility issue with the existing complementary metal-oxide-semiconductor (CMOS) manufacturing platform. Here, we demonstrate the occurrence of mid-infrared localized surface plasmon resonances (LSPR) in thin Si films hyperdoped with the known deep-level impurity tellurium. We show that the mid-infrared LSPR can be further enhanced and spectrally extended to the far-infrared range by fabricating two-dimensional arrays of micrometer-sized antennas in a Te-hyperdoped Si chip. Since Te-hyperdoped Si can also work as an infrared photodetector, we believe that our results will unlock the route toward the direct integration of plasmonic sensors with the one-chip CMOS platform, greatly advancing the possibility of mass manufacturing of high-performance plasmonic sensing systems.

Corresponding authors, email: [*]m.wang@hzdr.de; [‡]ye.yu@tu-dresden.de.

# I. Introduction

The field of plasmonics has elicited great attention and research efforts for potential applications as varied as enhanced sensing, waveguides for integrated optical interconnects, and nanoscale optoelectronic devices [1-3]. Surface plasmons generated by the coupling of collective charge oscillations with electromagnetic radiation at a conducting surface [4] offer effective light-matter interactions in nanoscale structures with sub-wavelength light confinement and large enhancement of the local electromagnetic field intensity [5,6]. In this framework, materials with a good crystalline quality and a small amount of plasmon losses are proposed for plasmonic applications that span a wide range of electromagnetic frequencies [5,7,8]. Particularly, the mid-infrared (MIR) spectral range that covers the frequency band from 3000 to 300 $cm^{-1}$ is of high interest for chemical and biological molecular sensing [9-11], since many molecules display unique spectral vibrational fingerprints in this range. Therefore, a plasmonic material platform that can be tailored to the frequency range of interest with strong subwavelength confinement is desired [12].

It has been recently shown that hyperdoped semiconductors [10,13-18] provide a promising material system for the development of a plasmonic platform with inherent advantages in the mid-infrared (MIR) region [19,20]. Here, hyperdoping means a doping concentration well above the impurity solid solubility limit. Contrary to metals (Au, Ag, and Al), in which the density of free electrons is fixed and the resulting resonance frequencies are mainly located in the visible and near-infrared (VIS-NIR) spectral range [11], the plasmon resonance in doped semiconductors can be tuned over a broad spectral window, ranging from NIR to far-infrared (FIR) by controlling the carrier concentration [12,15,21-26]. Among hyperdoped semiconductor materials, Si is the most desired material for plasmonic applications in the MIR spectral range, owing to its compatibility with the complementary metal-oxide-semiconductor (CMOS) technology [21,22,27-32]. Moreover, a plasmonic material based on doped Si has certain advantages over other doped semiconductors, such as cost-effectiveness, environmental friendliness and the well-developed and versatile fabrication process. Importantly, compared to III-V semiconductors, the absence of optical phonon absorption in the FIR spectral range in Si will naturally reduce the plasmon losses since Si is a non-polar semiconductor [15,33-35]. Recently, Si hyperdoped with the chalcogen dopant Te has gained increasing attention in the development of room-temperature MIR photodetectors [36]. Also, Te-hyperdoped Si exhibits a high free-electron concentration with a reasonably high mobility [37], as well as a better thermal stability as compared to other deep-level impurities [33]. This boosts the potential of Te-hyperdoped Si serving as a nanoscale MIR plasmonics platform with a broadly tunable plasma frequency. Such a platform could allow an easy integration with the current chip technology, providing room-temperature enhanced IR sensing.

In this work, we explore the potential of using Si hyperdoped with the deep-level dopant Te as an alternative for MIR-FIR plasmonic materials. A plasma frequency $\omega_p$ of around 1880 ~1630 cm$^{-1}$ is obtained in the Te-hyperdoped Si material. In addition, micrometer-sized antenna arrays, which are fabricated from the Te hyperdoped-Si layer via electron-beam lithography and reactive ion etching, exhibit an enhanced localized plasmon resonance in the spectral range of 100 to 700 cm$^{-1}$ compared to the non-patterned Te-hyperdoped Si material. This work points out the potential of Si hyperdoped with Te for plasmon-enhanced sensing applications such as the detection of the vibrational fingerprints of thin molecular films.

## II. Methods

### A. Experimental methods

A non-equilibrium approach combining ion implantation and flash lamp annealing (FLA) is used to achieve solid-phase epitaxial regrowth of hyperdoped-Si layers with a Te doping concentration several orders of magnitude above the equilibrium solid solubility limit. The electrical and optical properties of Si hyperdoped with Te have been systematically investigated previously [36,37]. It was found that the hyperdoped-Si layer with a Te concentration of 1.5% exhibits below-bandgap infrared absorption and yields high electron concentrations. In this work, we used (100)-oriented double-side polished Si wafers (*p*-type, boron-doped, $\rho \approx$ 1-10 Ωcm) with a thickness of around 380 μm. A triple implantation with energies of 350 keV, 150 keV and 50 keV with a fluence ratio of 5.3:2.3:1 was applied to compensate the Gaussian distribution of Te dopants during the implantation process [31]. The dopant concentration and the doping depth can be increased by increasing implantation fluence and energy, respectively. The parameters can be simulated by using the software of the Stopping and Range of Ions in Matter (SRIM) [38-40]. Note that, in our previous works [36,37], we used pulsed laser melting to recrystallize Te-implanted Si with a thickness of about 100 nm. Here, millisecond-range FLA is applied to recrystallize the implanted layer [41,42]. The FLA was performed in nitrogen ambient with a pulse length of 3 ms along with an energy density of around 58 J/cm$^2$. This corresponds to a temperature in the range of around 1200 ± 50 °C at the sample surface. Further details on the FLA system employed in our experiments can be found in the supplementary information (SI). The optimal FLA parameters were obtained by inspecting the crystalline quality of the implanted layers using micro-Raman spectroscopy.

The antenna patterning was achieved by electron-beam lithography and reactive ion etching (see SI for more details). The geometrical details of the two-dimensional Te-hyperdoped Si antenna arrays are schematically depicted in Fig. 1. The rectangular antenna arms have a size of 2 µm (length) × 0.8 µm (width), and two equal arms featuring different gap sizes varied between 0.2 and 1 µm.

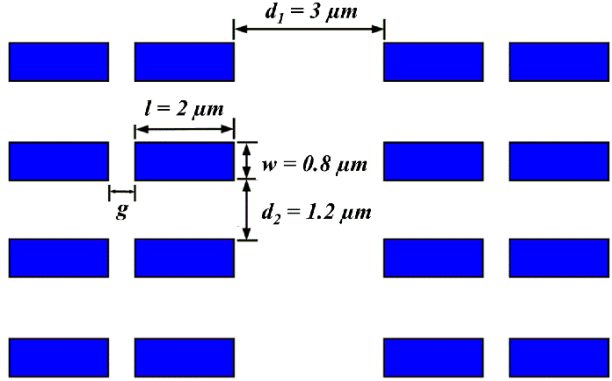

FIG. 1. A sketch of the geometry of the Te-hyperdoped Si antenna arrays. The geometrical details are indicated in the figure.

Structural characterization of the Te-hyperdoped Si layers and antenna arrays were performed by micro-Raman spectroscopy using a linearly polarized continuous 532 nm Nd:YAG laser for excitation and by scanning electron microscopy (SEM, S-4800, Hitachi). In addition, the micro-structural properties of FLA-treated Te-hyperdoped Si layer with a Te doping concentration of 1.5% were carried out by high-angle annular dark-field scanning transmission electron microscopy (HAADF-STEM) with energy-dispersive x-ray spectroscopy (EDXS) element distributions, further details concerning the TEM techniques employed in our experiments can be found elsewhere [40]. The electrical properties were characterized using a Lake Shore Hall measurement system in a van der Pauw configuration. IR measurements were carried out at room temperature by Fourier-transform infrared spectroscopy (FTIR, Bruker Vertex 80v). A Deuterated L-Alanine doped Triglycine Sulphate (DLaTGS) detector was used for the detection of MIR (from 10000 to 400 $cm^{-1}$) radiation, while a deuterated triglycine sulfate (DTGS) detector was used for FIR (from 700 to 10 $cm^{-1}$). FTIR measurements were performed in vacuum to eliminate the infrared absorption lines of the atmosphere. A reflection geometry was employed to quantify the absolute reflectivity of FLA-performed Te-hyperdoped Si layers and the antenna arrays. The reflectance of a gold mirror was measured as a 100% reflectance standard for the FTIR data.

B. Theoretical methods

Electromagnetic simulations were carried out using finite-difference time domain (FDTD) methods (FDTD 3D Electromagnetic Simulator, DEVICE Multiphysics Simulation Suite, Lumerical Inc. version 8.22.2072) [43].

Before FDTD simulations, the broadband optical properties of the material, more specifically the real and imaginary part of permittivity need to be determined first. The permittivity of bulk highly doped semiconductors can be modeled approximatively across a wide frequency range using the Drude model:

$$\varepsilon(\omega) = \varepsilon_s \left(1 - \frac{\omega_p^2}{\omega^2 + i\omega\Gamma}\right), \qquad \omega_p^2 = \frac{ne^2}{\varepsilon_s \varepsilon_0 m^*(n)} \quad (1)$$

$$\varepsilon(\omega) = \varepsilon' + i\varepsilon'' = \varepsilon_s \left(1 - \frac{\omega_p^2}{\omega^2 + \Gamma^2}\right) + i\varepsilon_s \left(\frac{\Gamma \omega_p^2/\omega}{\omega^2 + \Gamma^2}\right) \quad (2)$$

where the frequency-dependent permittivity $\varepsilon(\omega)$ is characterized by the screened plasma frequency $\omega_p$, the scattering rate $\Gamma$, the background dielectric constant $\varepsilon_s$ (for Si, $\varepsilon \approx 11.7$) and the permittivity of free space $\varepsilon_0$. The screened plasma frequency $\omega_p$ given above is characterized by the dielectric constant $\varepsilon_s \varepsilon_0$, the free-carrier density $n$, the charge of the electron $e$ and the non-parabolic doping-dependent effective mass of the free carriers $m^*(n)$ (0.27 $m_e$ for $n$-type heavily doped Si [26,44,45]). If $\Gamma$ is small, $\omega_p$ marks the approximate spectral position where the sign of the real component of the semiconductor permittivity ($\varepsilon'$) switches from positive to negative. Therefore, such kind of semiconductors can serve as an engineered MIR-FIR plasmonic material at frequencies (wavelengths, $\lambda_p = \frac{2\pi c}{\omega_p}$) lower (greater) than the plasma frequency of the material. Particularly, the plasma frequency can be accurately controlled by the doping concentration of the semiconductor.

### III. Results and discussion
#### A. Material properties of the Te-hyperdoped Si layer

Raman spectroscopy and Hall measurements were performed in order to characterize the structural (Fig. 2(a) and (b)) and electrical properties (Fig. 2(c) and -(d)), respectively (further data can be found in section D of the SI). The Raman spectrum of the as-implanted sample shows a very broad Raman band located at 460 cm$^{-1}$, which corresponds to a typical feature of the amorphous silicon layer created during the Te implantation process [46,47]. After FLA, only the sharp peaks at 520 cm$^{-1}$ and 303 cm$^{-1}$ are observed, which are ascribed to the transverse optical (TO) phonon mode and the second-order two transverse acoustic phonon (2TA$_{Si}$) scattering, respectively. This indicates that the crystalline structure of the FLA-treated Te-hyperdoped Si layer is completely restored with comparable quality as the virgin Si wafer. The cross-sectional HAADF-STEM image superimposed with the corresponding EDXS element distributions is shown in Fig. 2(b). Apart from the native oxide layer at the surface, Te is found to be quite evenly distributed within the top ~120 nm of the Si wafer. However, the Te dopants show segregation with further increasing depth. In our previous work, we used pulsed laser to melt Te-implanted Si, where the implanted layer is around 120 nm and the Te distribution is homogenous [36]. Here we use higher implantation energy and flash lamp annealing to fabricate thicker samples, leading to some segregation in the interface between in the implanted layer and the Si substrate. In order to get a more homogenous Te distribution in thicker layer, the parameters for flash lamp annealing (such as pulse duration and energy, pre-heating temperature and duration) should be optimized [48]. As shown in Fig. 2(c) and (d), the Te-hyperdoped Si displays $n$-type and metal-like conductivity. The sheet carrier density and the carrier mobility measured at 300 K are $1.5 \times 10^{15}$ cm$^{-2}$ and 32 cm$^2$/V·s, respectively (further data can be found in section D of the SI). As indicated in the cross-sectional TEM image (Fig. 2(b)), the Te impurities distribute within a depth of up to around 160 nm and the segregation occurs between 120-160 nm. Therefore, the estimated volume carrier density is from $1.25 \times 10^{20}$ to $9.4 \times 10^{19}$ cm$^{-3}$ depending on the assumed effective thickness. This indicates an electrical activation efficiency of around 15% at room temperature. According to the theoretical Drude formalism (as shown in Eq. (1) and (2)) and assuming our Te-hyperdoped Si layer as a bulk material, $\omega_p$ is calculated to be around 1880 ~1630 cm$^{-1}$, using the electrical transport data.

To cross-examine the accuracy of the plasma frequency of the bulk material obtained above with respect to the experimental values, we performed a further analysis of the reflection spectrum of the thin film using transfer matrix method (TMM). Figure 2(c) displays the normal-incident reflectance ($R/R_{Au}$) spectra of the Te-hyperdoped Si layers measured by FTIR at room temperature. The spectrum of a Si wafer was also measured for comparison. The red dash-dotted line displays a three-layer system simulation of the reflectivity spectrum of Te-hyperdoped Si as described in the following. The approximated dielectric function of the implanted layer was constructed based on the multi-layer transfer matrix formalism with 20° incidence angle via VASE (vase-ellipsometer-liquid-prism-cell) software [49]. This model consists of two layers: (i) the implanted layer described by the Drude approximation function, and (ii) the Si substrate described by the tabulated optical constants of monocrystalline Si. The fitting gives a carrier concentration of $1.27 \times 10^{20}$ cm$^{-3}$ and plasma frequency is around 1900 cm$^{-1}$ with the input free electron mass of 0.27 $m_e$, which is reasonable for $n$-type deep-level impurity heavily doped Si [26,44,45]. The extracted plasma frequency from the fitting is shifted to a slightly higher wavenumber compared to the one obtained from the Drude formalism. The discrepancy may be related to the rather thin film thickness together with the graded doping; as a consequence, the fit allows many different solutions of similar quality, yet none could be found with optimal agreement. Besides, the line shape of the spectrum can be affected by the measurement conditions, such as the brilliance of the light source [50,51], the maximum angle of acceptance of the detector [52]. One must carefully examine the experimental and instrumental conditions used during the measurement of the reflectance spectra, from which the broadband permittivity of the material is deducted. Fig. 2(d) displays the temperature-dependent spectral reflectance of the Te-hyperdoped layer. The reflectance spectrum at different temperatures exhibits similar behavior, which suggests that Te-hyperdoped Si can be used as a plasmonic material in a broad temperature range. In addition, the plasma frequency and optical response of Te-hyperdoped Si can be tuned by altering the doping concentration as well as the electrical and optical excitation.

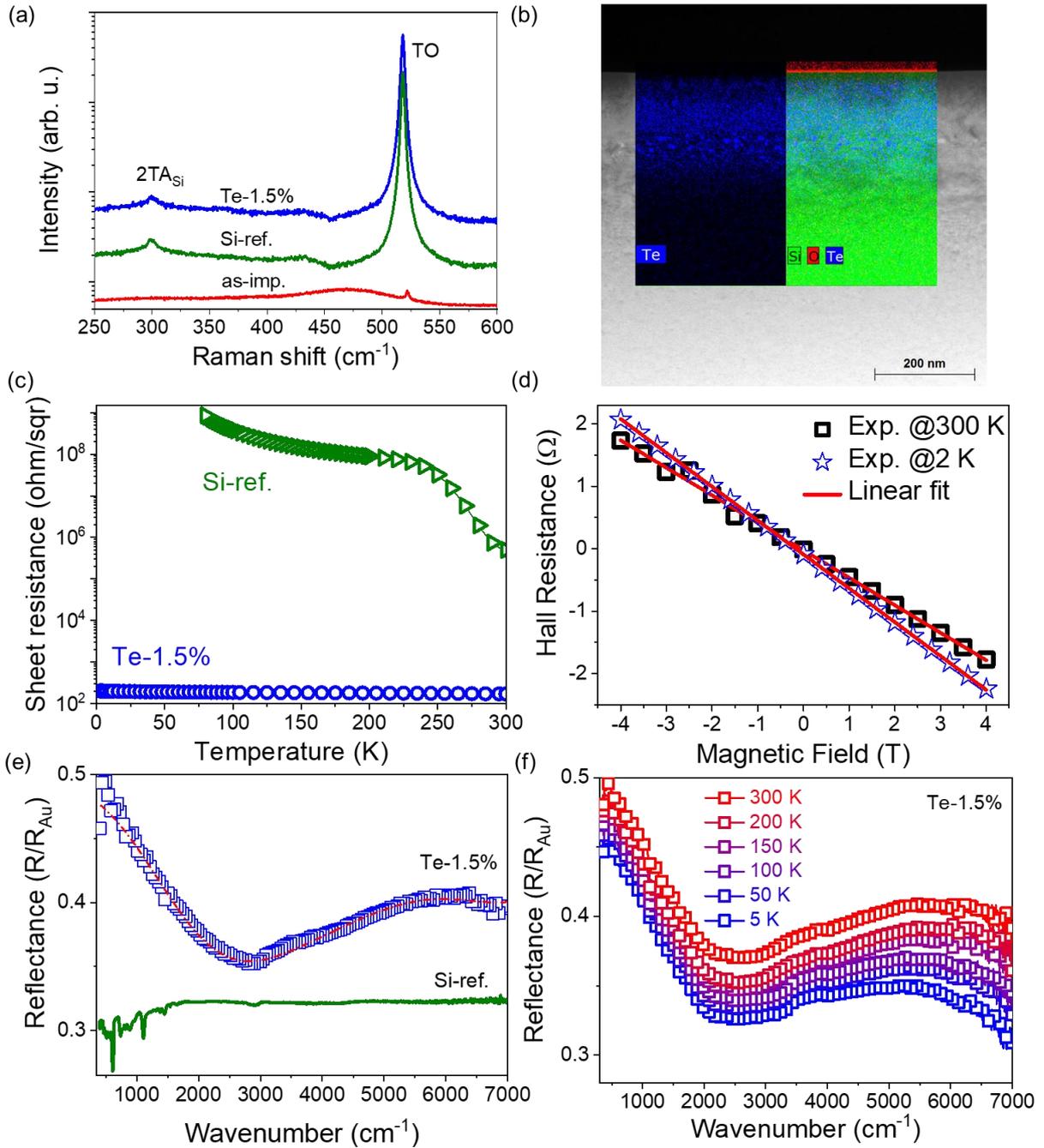

FIG. 2. Material characterization for FLA-treated Te-hyperdoped Si sample with the doping concentration of 1.5%. (a) Room-temperature micro-Raman spectra. The Raman spectra of the as-implanted sample as well as virgin Si (red and green solid lines, respectively) are also included for comparison. The spectra are stacked vertically with a constant offset for clarity. (b) Cross-sectional HAADF-STEM image superimposed with the corresponding EDXS element distributions (blue: tellurium, green: silicon, red: oxygen). Apart from the native oxide layer at the surface, Te is found to be quite evenly distributed within the top ~120 nm of the Si wafer. However, the Te dopants show segregation with further increasing depth. (c) The temperature-dependent sheet resistances. Also, the Si substrate is included as comparison. (d) The Hall resistance at 2 K and 300 K with a linear fit to experimental data. (e) Room-temperature reflectance spectra. The reflectivity of a bare Si wafer is included as a reference. (f) Reflectance measurements at different temperatures.

**B. LSPR in Te-hyperdoped Si: Electromagnetic simulations**

In order to understand and design the antenna response, we have performed FDTD simulations. We started with the simplest structure to simulate the spectral response of one nanobar with an atomic Te doping concentration of 1.5% on a Si substrate in order to first analyze the possible range of plasmonic resonances. The dimensions of the antennas were set to be the same width ($w$) of 0.8 μm and thickness of 0.3 μm, as mentioned in the previous section, except the arm length ($l$) was varied from 0.5 to 3 μm. The corresponding extinction spectra and electric field intensity distribution are shown in Fig. 3. As displayed in Fig. 3(a), when the polarization is perpendicular to the long axis of the antenna, the spectral position of the resonance does not change as the antenna length increases. On the other hand, when the electric field polarization is parallel to the long antenna axis, the plasmon resonance shows a red shift (for a depiction in linear wavelength scale see Supporting Information Fig. S1) with increasing antenna length (Fig. 3(b)). The difference in the extinction cross-section shown in Fig. 3 (a) and (b) indicates that the LSPR mode in Fig. 3(a) is much less intensive than that in Fig. 3(b). To understand the nature of these two modes, we further calculated the corresponding electric field intensity distributions at the resonance positions (at the wavelength of 25 μm), which is displayed in the top-down view in Fig. 3 (c) and (d), respectively. As displayed in the coordinate, the antenna has a width of 0.8 μm and an arm length of 2 μm. As the polarization direction is perpendicular to the long axis of the antenna, shown in Fig. 3(c), the free electrons only accumulate along the longer edge (length), indicating that this plasmonic resonance is a transverse mode. Whereas, in the case of Fig. 3(d), since the polarization direction is parallel to the long axis of the antenna, the free electrons in turn accumulate along the shorter edge (width), exciting the longitudinal mode at longer wavelength (Fig. 3(d)). This in general results in the strong near-field enhancement, which will be beneficial in surface-enhanced spectroscopies.

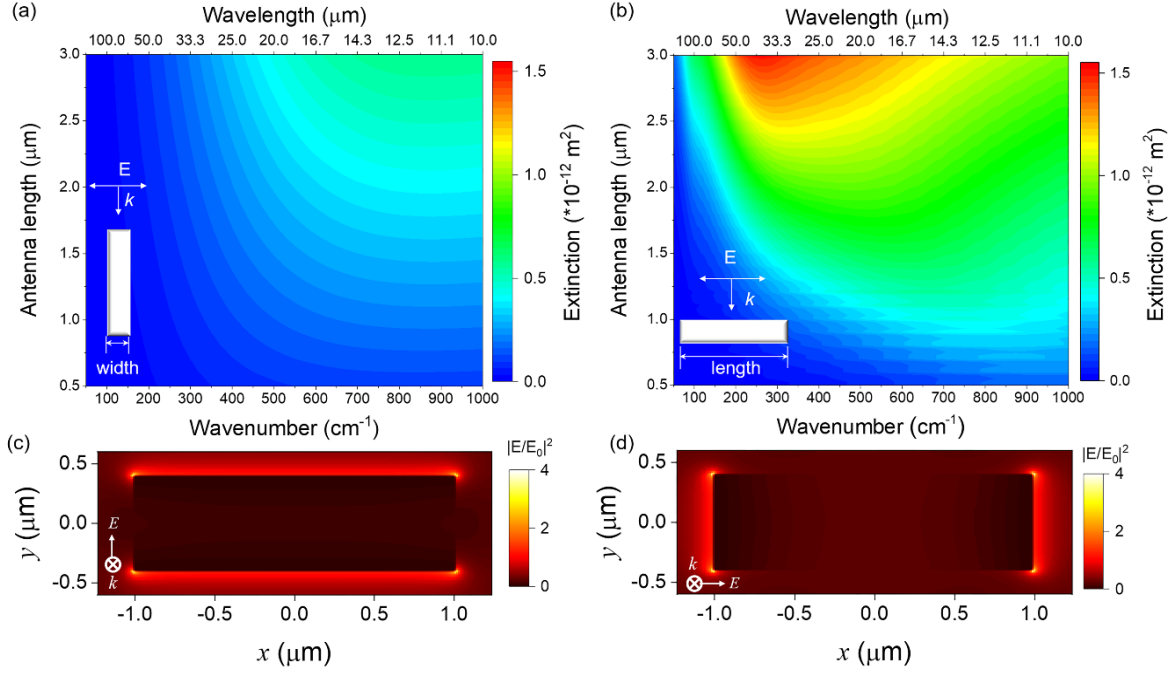

FIG. 3. The simulated extinction spectra as a function of the antenna length of a single antenna with the electric field polarization perpendicular to the long antenna axis (a) and parallel to the long antenna axis (b). The antenna length against wavelength is plotted in Fig. S1 in the Supporting Information for clarity. Top-down view of the electric field distribution map of single-arm antenna at the LSPR position (at the wavelength of 25 µm), the electric field polarization perpendicular to the long antenna axis (c) and parallel to the long antenna axis (d). The electric field enhancement is obtained by normalizing the electric field intensity to the incident plane wave amplitude $E_0$.

From the FDTD simulations of the single antenna in Fig. 3, we have successfully obtained the spectral range of the LSPRs for such a system. We then investigate the plasmonic properties of an array of coupled antennas, *i.e.* two nanoantennas placed very close to each other, forming a repeating structural unit, as illustrated in Fig. 1. The separations between units along the *x* and *y* axis were set to be 3 µm and 1.2 µm, respectively. The dimensions of the nanoantennas were kept constant: $l = 2$ µm; $w = 0.8$ µm; and $t = 0.3$ µm. Only the gap size between the two adjacent antennas was varied from 0.1 µm to 1 µm. However, since both periodicities are much smaller than the wavelength of the incident radiation resonant to the plasmon of Te-hyperdoped Si, we expect the diffractive far-field coupling not to interfere with the LSPRs. Fig. 4(a) reveals the broad-band spectral response of the coupled system, showing two reflection peaks sandwiching a reflection dip in between, indicating that the coupling between two adjacent nanoantennas is present. The presence of coupling is also corroborated by the fact that, as the gap size increases, the reflection peaks experience a slight blue shift due to the decreased near-field coupling strength. However, further careful examination is still needed, since the reflection intensity is expected to be consistent with the transmission intensity. Hence, the electric field intensity distribution, along with the corresponding charge distribution of the nanonantenna pair at both reflection peaks and valleys were plotted in the cross-section view, shown in Fig. 4(c) through (f), taking the structure with gap size $g = 0.2$ µm as an example. The spectral positions were all marked correspondingly in Fig. 4(b). Simulation reveals that at small wavenumbers, the system exhibits predominately the characteristics of a coupled dipolar resonance [53,54], such as Fig. 4(c) and (d). This indicates that at a longer wavelength, an additional mode is created due to the coupling between the two adjacent nanoantennas. However, at shorter wavelength, the system exhibits briefly a quadrupolar resonance, shown in Fig. 4(e), and then transitions into non-coupled dipolar resonance (Fig. 4(f)). Note that the six-node charge distribution in Fig. 4(e) is because the energy level of the quadrupolar resonance is too close to the non-coupled dipolar resonance. This plasmonic behavior over the spectrum of interest suggests a classic plasmonic hybridization model, where the longitudinal dipolar resonances are treated as electric dipoles and the quadrupolar resonances as an electric dipole pair with opposite moment, the sketch of which is shown in Fig. S2 in Supporting Information.

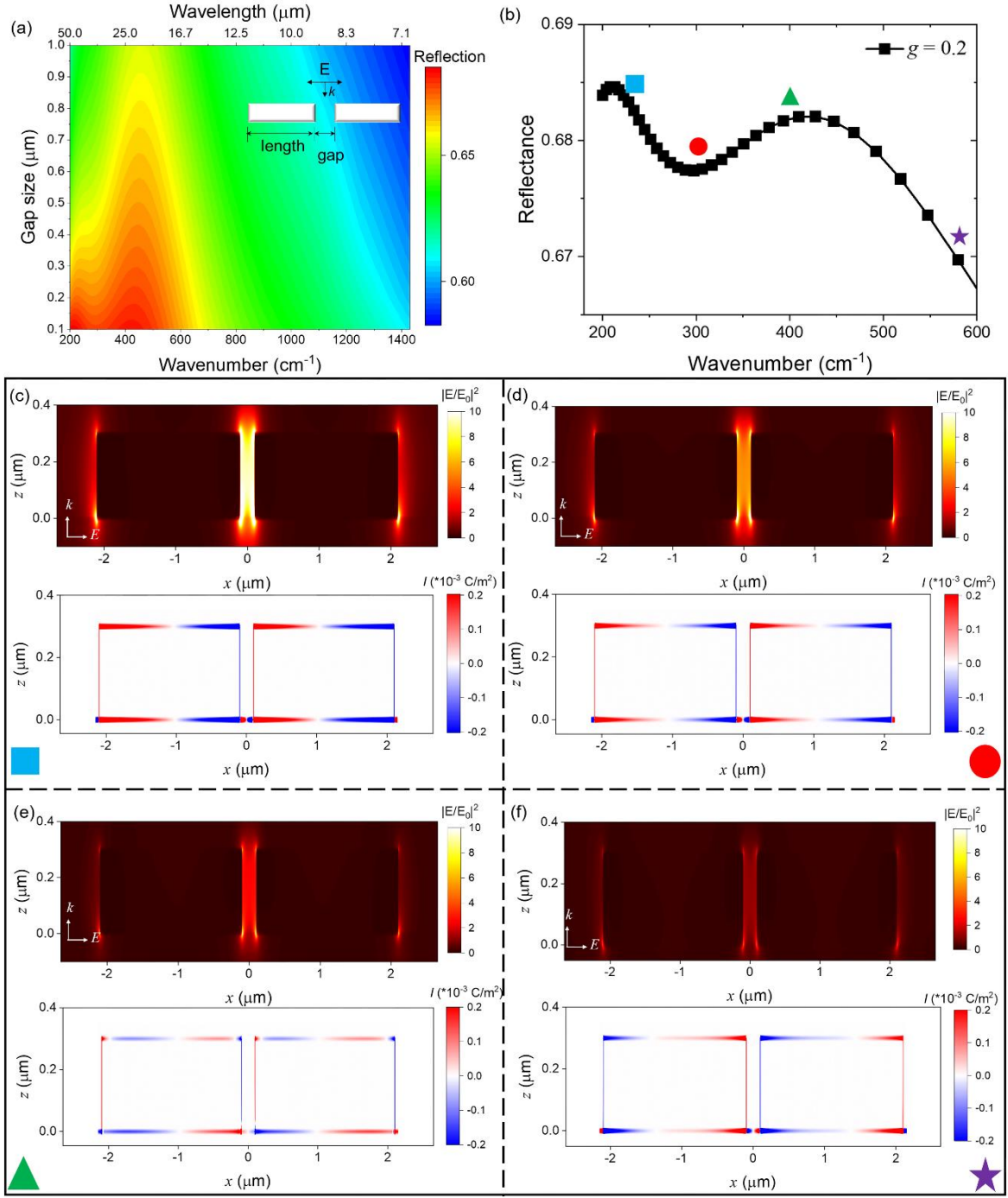

FIG. 4. The FDTD simulations for the paired-armed antennas. The simulated antenna has a width of 0.8 μm, a thickness of 0.3 μm and an arm length of 2 μm. (a) The simulated reflection spectra as a function of the gap size in an array of paired-arm antennas (antenna arm length, 2 μm) with the electric field polarization parallel to the long antenna axis. (b) - (f) are simulation results of a nanoantenna pair consisting of two identical nanobars with a gap size of 0.2 μm. (b) Simulated reflection spectrum and the cross section view of the electric field enhancement distribution (upper panels in c-f), as well as the charge distribution (lower panels in c-f) at the wavelengths of 44 μm (227 cm$^{-1}$) (c, marked with a blue solid square), 33 μm (303 cm$^{-1}$) (d, marked with a red circle), 25 μm (400 cm$^{-1}$) (e, marked with a green triangle), and 17 μm (588 cm$^{-1}$) (f, marked with a purple star), respectively. The electric field enhancement is obtained by normalizing the electric field intensity to the incident plane wave amplitude $E_0$.

Upon plasmonic hybridization, the dipolar resonances form a bonding mode, occupying a lower energy level, as well as an anti-bonding mode, occupying a higher energy level. At the same time, the quadrupolar resonances experience the same process, creating two additional energy levels. However, from FDTD simulations, only two energy levels were observed (Fig. 4(c) - (e)), indicating that the two anti-bonding modes are likely residing at the energy levels beyond the plasma frequency. On the other hand, we also observed that the non-coupled dipolar resonance remains prominent over a broad range of the spectrum, as evidenced in Fig. 4(f). This result indicates that even with a relatively small gap size that we simulated, the coupling strength is relatively low, which is corroborated by the small permittivity of the Te-hyperdoped Si material, as compared to the conventional plasmonic materials such as gold. It is safe to assume that, if a system with high coupling strength is needed, a much smaller gap size is generally required. However, this does not negate the value of plasmonic semiconductor materials. First, the "weak" confinement, despite the lower local field intensity, allows the plasmon resonances to reach the bulk environment with large interrogation volume. This could provide a more stable sensing performance cross long distances from the semiconductor/dielectric interface, as well as potentially lower the detection limit. Second, the propagation loss from the plasmonic semiconductor material is very low, since the imaginary part of the permittivity is a few orders of magnitude lower than that of plasmonic metals. While this is by no means the defining factor of a "better" material, it does provide a unique feature as compared to classic plasmonic metals. Additionally, the Te-hyperdoped Si material can result in an effective $n$-type doping with carrier density approaching $10^{21}$ cm$^{-3}$ [37], which is only one order of magnitude lower than that of metal, in conjunction with a few comparative studies [55,56], it would be reasonable to assume that the plasmonic performance from the Te-hyperdoped Si material could be on par with plasmonic metal structures.

**C. LSPR in Te-hyperdoped Si: Infrared spectral properties**

The hyperdoped Si paired-arm antenna arrays with a Te doping concentration of 1.5% were designed based on the considerations of providing maximum coupling between the paired antenna arms and preserving the maximum coverage of the substrate. Therefore, the period for the antenna arrays is designed to be large enough to avoid significant near-field interactions between two neighboring paired antennas, where the near field coupling is only possible between the two nanobars within the same pair, as verified by the electromagnetic simulations (FDTD electric field intensity map) shown in Fig. 4. At the same time, the period for the antenna arrays should be smaller than the wavelength of the incident radiation resonant to the plasmon to avoid crosstalk between LSPR and diffractive far-field coupling. Thereby, with a small period, we can ensure that the spectral response is strong enough since more nanostructures will be measured in the same unit area. In this design, we can ensure that the plasmonic antennas provide spectral response strong enough to be detected. Moreover, the gap (*g*) between the two arms is varied from 0.2 μm to 1 μm, which aims at affecting the local intensity distribution, thereby further boosting the enhancement and confinement of the field. Representative SEM images of Te-hyperdoped Si paired-arm antenna arrays are presented in Fig. 5, where the antennas have a width (*w*) of 0.8 μm, an arm length (*l*) of 2 μm and consist of two equal arms featuring a gap size of 0.2 μm (Fig. 5(a)), 0.3 μm (Fig. 5(b)), 0.4 μm (Fig. 5(c)), and 1.0 μm (Fig. 5(d)). In order to investigate the sensing applications of Te-hyperdoped Si antenna arrays, it is worth discussing the antenna signal in reflection geometry. Particularly, it is important to use the reflectance spectrum in realistic MIR chemical sensing devices, since the aqueous solutions are not transparent at MIR frequencies.

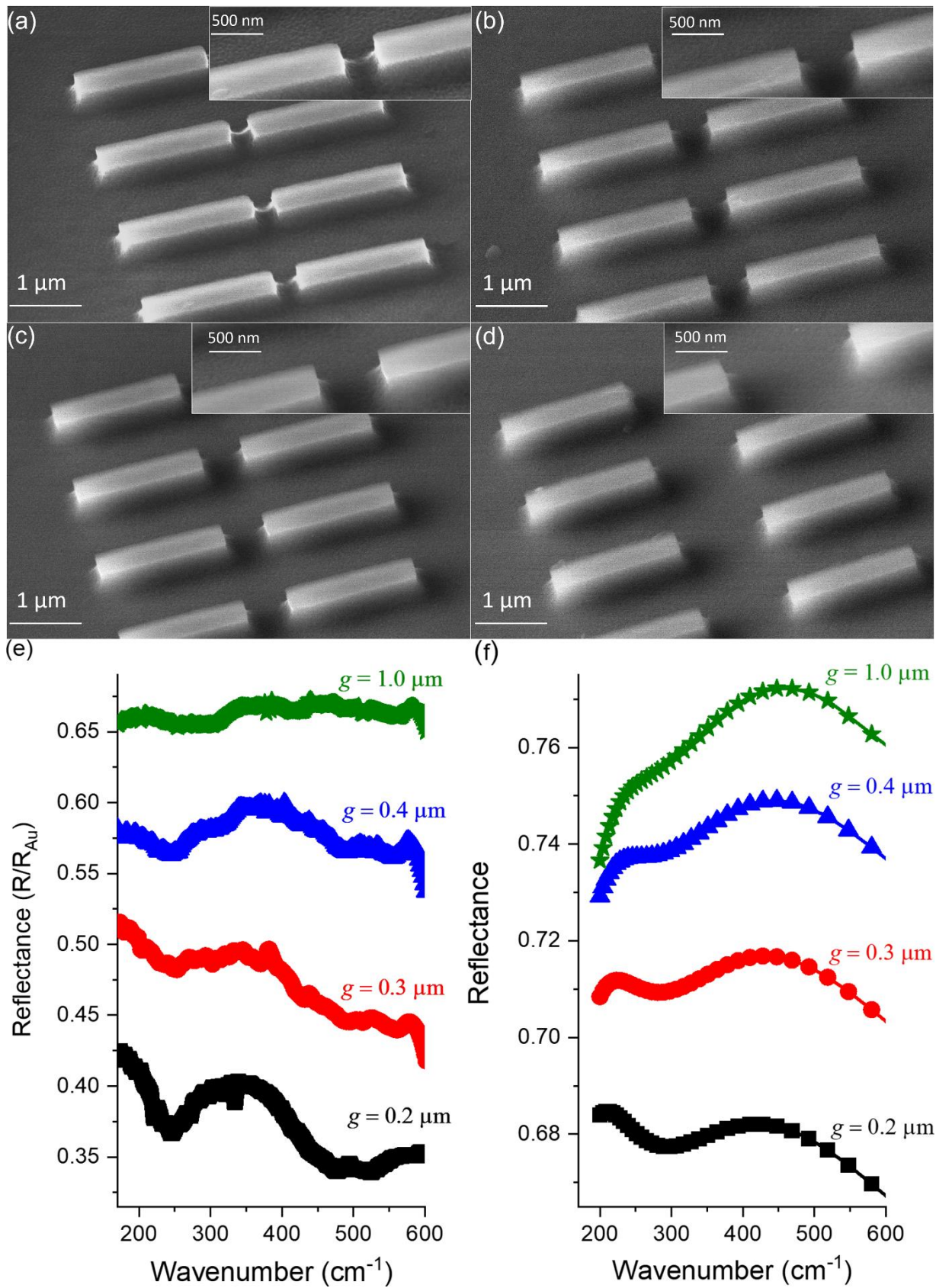

FIG. 5 Demonstration of MIR-FIR localized surface plasmon resonances in Te-hyperdoped Si antenna arrays. Bird´s-eye SEM images of antenna arrays with an antenna width of 0.8 μm, an arm length of 2 μm and two equal arms featuring different gap sizes: (a) 0.2 μm, (b) 0.3 μm, (c) 0.4 μm and (d) 1.0 μm. (e) Representative reflection

spectra for the antenna arrays shown in (a)-(d). The spectra are measured with the electric field polarization parallel to the long antenna axis. (f) The simulated reflection spectra of the same antenna arrays in (e), which correspond to horizontal cuts through Fig. 4 (a). The spectra in (e) and (f) have been vertically offset with a constant of 0.1 for clarity.

Figure 5(e) demonstrates the reflection spectra in the far-infrared range at normal incidence with the electric field polarization direction parallel to the long antenna axis. The spectra recorded in the mid-infrared range are shown in Fig. S3 in the SI. We conducted a comparison of the measured reflection spectra of the paired-arm antenna arrays with four different gap size of 0.2, 0.3, 0.4, and 1.0 μm, with $l$ = 2.0 μm and $w$ = 0.8 μm for an individual antenna arm. In order to extract the antenna optical response, the reflectance is displayed as $R/R_{Au}$, with $R$ and $R_{Au}$ being the reflection spectra acquired from the antenna array sample and from a gold mirror, respectively. The experimental results clearly display two sets of reflection peaks and valleys, shown in Fig. 5(e). In addition, these features barely observed when the excitation electric field is oriented perpendicular to the long antenna axis (shown in Fig. S4 in SI). This suggests that the line shape that we observe in Fig. 5(e) is indeed caused by the presence of longitudinal plasmonic modes. Furthermore, a blue shift of the features with increasing gap size is observed, indicating that the coupling between two adjacent nanobars is present. Figure 5(f) presents the simulated reflection spectra for the same antenna arrays shown in Fig. 5(e). The line shapes of the two sets of spectra show reasonable resemblance to each other, indicating that our simulation can provide a good insight to the plasmonic behavior of the system. However, discrepancies are also noticeable. We believe, they mainly originate from the inhomogeneity of the Drude response, because the Te doping profile is not perfectly uniform along the depth, as shown in Fig. S5 in SI. One important aspect that should be noted, is that for plasmonic semiconductors, an accurate broadband permittivity ($\varepsilon(\lambda)$ and $\varepsilon'(\lambda)$) is the key to the quality of simulation, and has to be extracted properly. In our simulations, the permittivity of the Te-hyperdoped Si material was extracted from the classic Drude mode calculation, which only represents the theoretical values for the bulk material. In reality, the permittivity of the same material but in thin-film configuration often varies, which is partially evidenced by the discrepancies between the Drude model and the fitted data obtained via the TMM shown in Fig. 2. Moreover, the plasmon damping near the edges of the antenna structures may be much higher than in the inner part due to additional scattering by the roughly etched surfaces and nonideally shaped cuboid (as shown in the insets of Fig. 5(a)-(d)), which can be further minimized by optimizing the etching parameters.

## IV. Conclusions

In summary, we have demonstrated that Si hyperdoped with the deep-level impurity Te is a promising material platform for MIR-FIR plasmonic applications. We have accomplished electron density of about $1.25\times 10^{20}$ to $9.4\times 10^{19}$ cm$^{-3}$ in Te-hyperdoped Si, which gives rise to a plasma frequency $\omega_p$ of around 1880 ~1630 cm$^{-1}$ ($\lambda_p = 5.3$~$6.1$ μm), ideal for surface-enhanced infrared absorption spectroscopy. By fabricating two-dimensional antenna arrays out of the Te-hyperdoped Si, multiple plasmon resonance modes are excited in the frequency range of 100-600 cm$^{-1}$. Since the whole manufacturing process of the Te-hyperdoped Si antennas is compatible with the CMOS technology, our result holds great promise for the realization of CMOS-compatible MIR-FIR devices for substance-specific molecular sensing, IR imaging, light detection, and energy harvesting.

## Conflicts of interest

There are no conflicts to declare.

## Acknowledgements

Authors acknowledge the ion implantation group at HZDR for performing the Te implantation. Additionally, support by the Structural Characterization Facilities at the Ion Beam Center (IBC) and SEM characterization by Elfi Christalle are gratefully acknowledged. M.W. and M. S. S. thank the financial support by the Deutsche Forschungsgemeinschaft (DFG) (grant No. WA4804/1-1). M.W. thanks Claudia Neisser and Tommy Schönherr for their assistance with the electron beam lithography experiments, and Denny Lang for his assistance with the FTIR measurements. Y.Y. thanks the support by the DFG under the project 419981939 (PlasCode). R. K. is supported by the DFG through the "Heisenberg Program" (Project Number 326062881). The authors would like to thank the "Research Cloud Services" from the Center for Information Services and High Performance Computing (ZIH) of Technische Universität Dresden for their support in the computational resources.